\documentclass[preprint,authoryear,12pt]{elsarticle}

\usepackage{epsfig}

\usepackage{amssymb}
\usepackage[hyphens]{url}

\journal{Advances in Space Research}
\def\arcsec{\hbox{$^{\prime\prime}$}}

\def\degns{\ifmmode^\circ\else$^\circ$\fi}
\def\deg{\ifmmode^\circ\else$^\circ$\fi}

\begin{document}

\begin{frontmatter}



\title{First solar observations with ALMA}


\author{Maria Loukitcheva\corref{cor}}
\address{Saint Petersburg State University, 7/9 Universitetskaya nab., St. Petersburg 199034, Russia}
\address{St. Petersburg branch of Special Astrophysical Observatory, Pulkovskoye chaussee 65/1, St. Petersburg 196140, Russia}
\cortext[cor]{Corresponding author}
\ead{lukicheva@mps.mpg.de; m.lukicheva@spbu.ru}


%
%

\begin{abstract}

The Atacama Large Millimeter-Submillimeter Array (ALMA) has opened a new window for studying the Sun via high-resolution high-sensitivity imaging at millimeter wavelengths. In this contribution I  review the capabilities of the instrument for solar observing and describe the extensive effort taken to bring the possibility of solar observing with ALMA to the scientific community. The first solar ALMA observations were carried out during 2014  and 2015 in two ALMA bands, Band 3 ($\lambda=3$~mm) and Band 6 ($\lambda=1.3$~mm), in single-dish and interferometric modes, using single pointing and mosaicing observing techniques, with spatial resolution up to  $\sim$2\arcsec\ and $\sim$1\arcsec\ in the two bands, respectively. I overview several recently published studies which made use of the first solar ALMA observations, describe current status of solar observing with ALMA and briefly discuss the future capabilities of the instrument.
%

\end{abstract}

\begin{keyword}
Solar atmosphere; Solar chromosphere; Millimeter radiation; ALMA
\end{keyword}

\end{frontmatter}

\parindent=0.5 cm


\section{Introduction to ALMA}
\begin{figure}
\label{fig1}
\begin{center}
\includegraphics*[width=11cm]{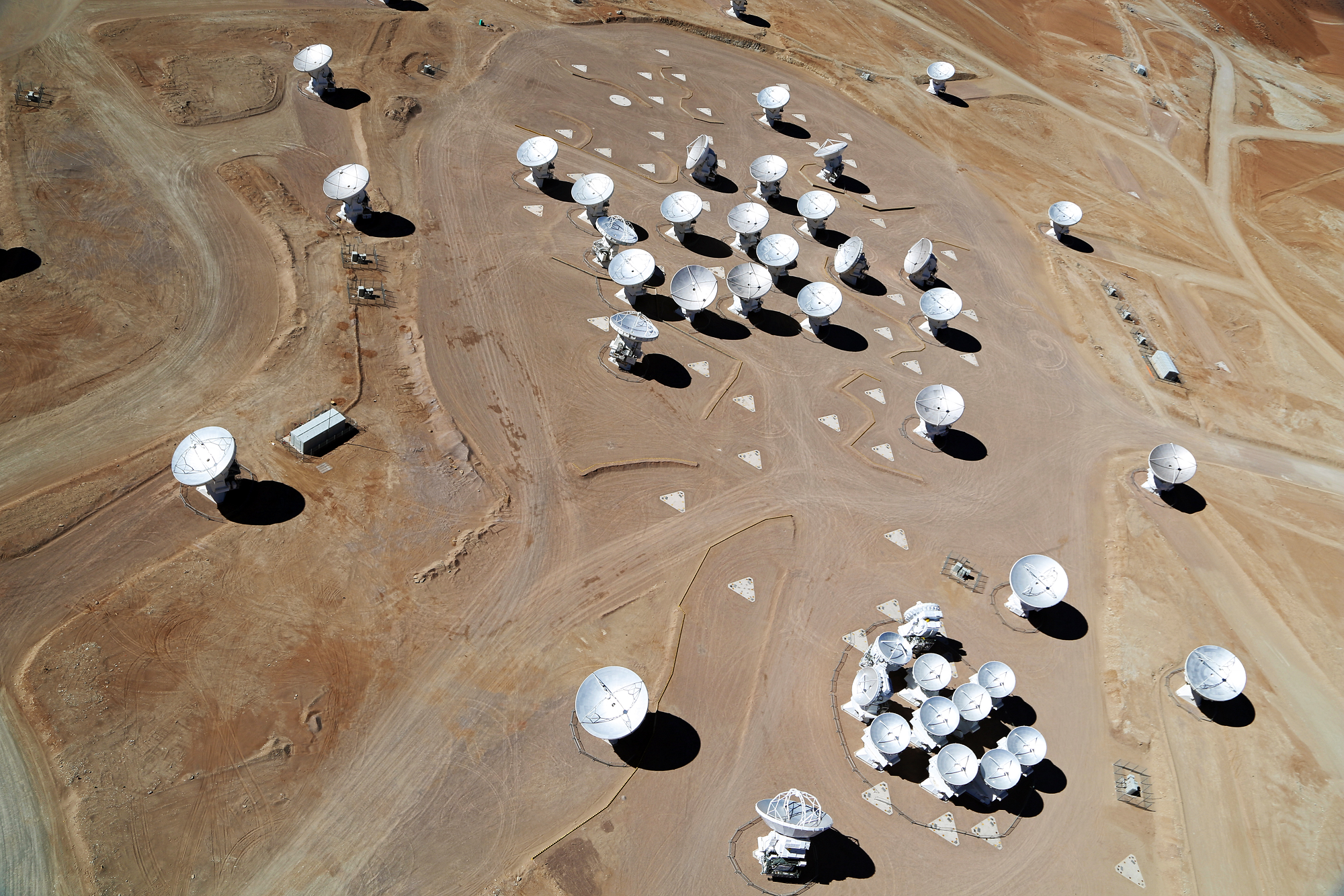}
\end{center}
\caption{The aerial image of the ALMA interferometer captures the compact array and a part of the 12-meter array. Credit: Clem \& Adri Bacri-Normier (wingsforscience.com)/ESO.}
\end{figure}

The Atacama Large Millimeter/Submillimeter Array\footnote{\url{http://www.almaobservatory.org/}} (ALMA) is the largest ground-based astronomical project in existence. ALMA is located at an altitude of 5000~m in the Chilean Atacama dessert and is operated in a world-wide international cooperation by the European Organization for Astronomical Research in the Southern Hemisphere (ESO), the National Radio Astronomy Observatory (NRAO) and the National Astronomical Observatory of Japan (NAOJ).  ALMA is a radio interferometer composed of 66 high-precision antennas: fifty 12-meter antennas for  high-resolution high-sensitivity imaging (the 12m Array), twelve 7-meter antennas and four 12-meter total power antennas, forming together the Atacama Compact Array to enhance wide-field imaging \citep[][Figure~\ref{fig1}]{Wootten,Hills}.

ALMA receivers cover the range of wavelengths from $\sim$0.3~mm (corresponding frequency of 950~GHz) to $\sim$3.6~mm (84~GHz) with prospects of going to longer wavelengths up to 9 mm \citep{Huang2016}. Depending on the required flux sensitivity and/or spatial resolution the interferometer can be used in a number of different antenna configurations: from compact (with maximum baselines as small as 150~m) to extended (with baselines up to 16 km). The spatial resolution depends on the array configuration, in particular on the maximum baseline, and wavelength. For the most compact configurations the spatial resolution reaches 1\arcsec\ at $\lambda=3$~mm and 0.5\arcsec\ at $\lambda=1$~mm, while for the longest baseline of 16 km it can be as high as 0.05\arcsec\ at $\lambda=3$~mm and  0.005\arcsec\ at the shortest wavelength of 0.3~mm. 

The instantaneous field of view (FOV) of the array depends on the primary beam size of the array antennas and wavelength, yielding ALMA FOV size  $\sim$60\arcsec\ when observing at $\lambda=3$~mm,  and $\sim$25\arcsec\ for $\lambda=1$~mm. For programs that require larger FOV, mosaicing is offered.  For instance, a 150-pointings mosaic at 3~mm will map a region on the Sun of size 350\arcsec x350\arcsec, while at 1~mm the size of the mapped region will be twice smaller.

\section{Solar science with ALMA}
The importance of ALMA for solar physics is indisputable. Solar continua at mm and submm wavelengths originate from the chromosphere, the most enigmatic solar atmospheric layer, which plays a significant role in defining the dynamics, energy and mass budgets of the corona and solar wind. The opacity of the gas that is sampled by ALMA increases with wavelength, by observing at different  wavelengths we probe different chromospheric layers/structures. 
With ALMA frequency bands we can cover a substantial part of the chromosphere from the temperature minimum up to the transition region to corona. Formation of mm/submm continua is well understood \citep[e.g.][]{dulk}: the radiation is produced by thermal free-free mechanism with the opacity due to encounters between free electrons and protons  ($H^0$ opacity), and opacity due to encounters between free electrons and neutral hydrogen atoms ($H^-$ opacity). The radiation is formed under the conditions of the local thermodynamic equilibrium (LTE) with the Planck function as a source function. As mm/submm wavelengths lie in the Rayleigh-Jeans regime, the observed intensity (brightness temperature) at a given wavelength is linearly proportional to the temperature of the emitting material.
This unique capability of linearly sampling the temperature of the optically-thick chromospheric gas with ALMA provides a big advantage over chromospheric diagnostic using optical spectral lines, which are formed under the non-LTE conditions and observations in these lines are not straightforward to interprete. 

Given the diagnostic potential of the mm-$\lambda$ regime, an extremely broad program of science in both active and quiescent Sun is possible with ALMA observations.
The potential science cases for ALMA are comprehensively discussed in \citet{Wedemeyer2016}.  ALMA will contribute into the studies of chromospheric thermal and magnetic structure, of chromospheric heating and dynamics determining the role played by waves, shocks, and magnetic reconnection; into the insights about the fine structure of filaments, prominences, and spicules; into study of the chromospheric reaction to coronal flares and chromospheric mictoflares; and into detection of solar line emission: recombination lines, Zeeman effect, molecular lines. 

The observational data, obtained during the commissioning phase and introduced in the ALMA reference papers, were released to the scientific community in January 2017\footnote{The science verification data is available for download at \url{https://almascience.nrao.edu/alma-data/science-verification}\label{sv}}. The available data sets consist of calibrated uv-data and scripts to synthesize images, and include both single-pointing and mosaic observations of quiet-Sun regions (QS) and active regions (AR). 
Although the number of ALMA antennas used in the science verification campaign did not exceed 36 (out of 66 ALMA antennas), and the spatial resolution did not exceed 2\arcsec\ at 3~mm and 1\arcsec\ at 1.3~mm, respectively, the solar images synthesized from the ALMA commissioning data were extensively used for various studies and a number of scientific papers were published, providing new insights about the solar chromosphere.

\begin{figure}
\label{fig2}
\begin{center}
\includegraphics*[width=12cm]{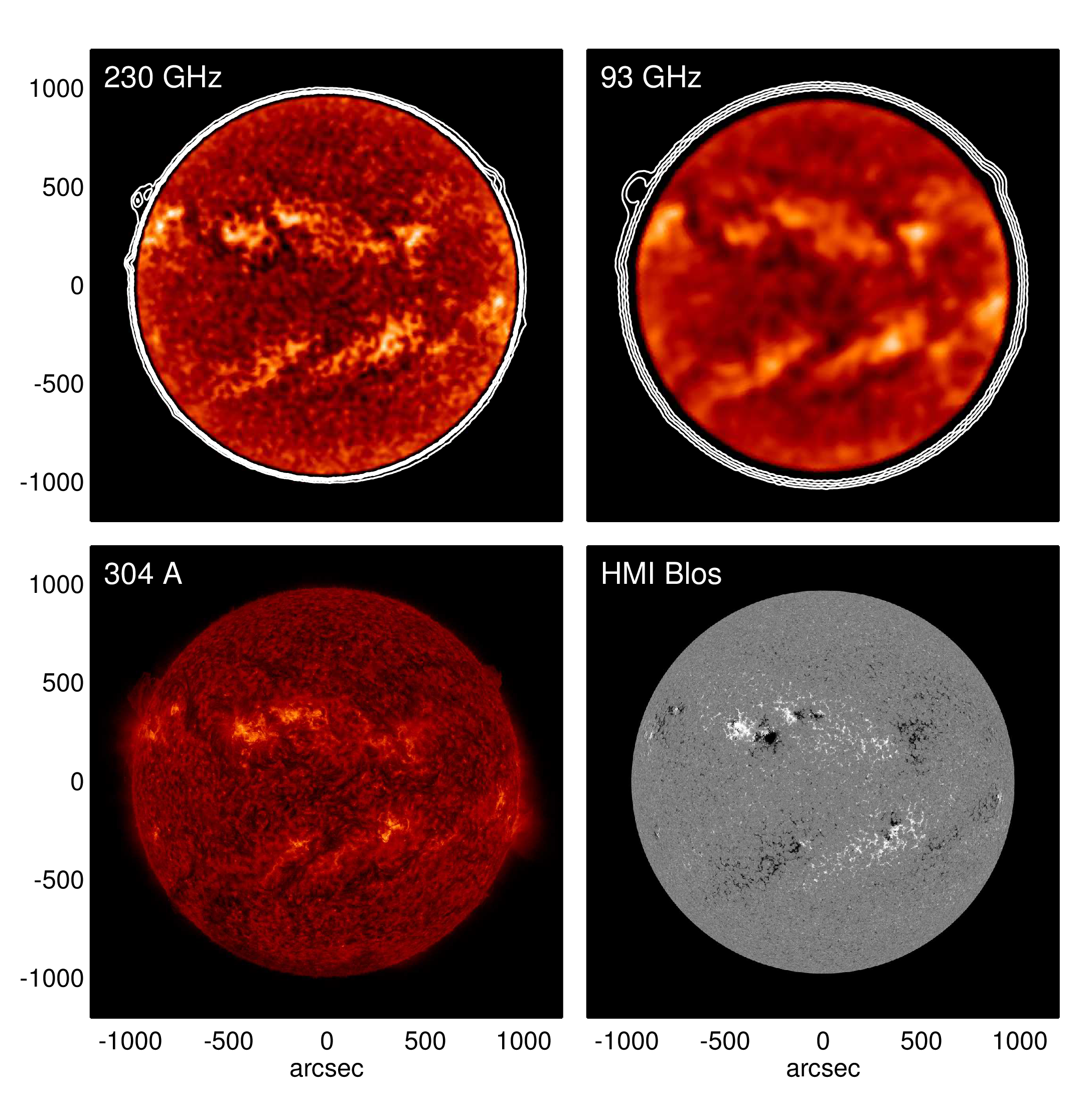}
\end{center}
\caption{Comparison of ALMA fast-scan TP maps obtained on 17 December 2015 at $\lambda=1.3$~mm (upper-left
panel) and $\lambda=3$~mm (upper-right panel) wth Solar Dynamics Observatory (SDO) Atmospheric Imaging Assembly (AIA) image at 304\AA\ (lower-left panel), and a line-of-sight magnetogram from the Helioseismic and Magnetic Imager (HMI) on board SDO (lower-right panel). The color displays range from 5300 to 7400 K for the image at 1.3~mm, and from 6700 to 8800~K for 3~mm. Low-level contours are added at 300, 600, 1200, and 2400~K to show features above the limb. Figure from \citet{White2017a}. Reprinted with permission from \textcopyright Springer Nature.}
\end{figure}

\section{Preparing for solar ALMA}

 Although ALMA is a general purpose instrument, the possibility of solar observations was incorporated in the array design from the very beginning \citep{Bastian2002}. To this end, the surface of the antennas was designed to scatter the optical and infrared radiation and to prevent the telescope hardware from overheat. But unlike other mm sources,  the Sun is a very extended object and it fills the FOV with low-contrast brightness structures varying on many temporal and spatial scales. The Sun is significantly brighter (5000--8000~K at mm wavelengths) than the typical temperatures ALMA receivers were designed and tested for (up to $\sim$800~K), which implies that the solar flux needs to be attenuated or the receiver gain must be reduced. In addition, the Sun possesses differential rotation, which should be tracked when acquiring both mosaic and sit-and-stare observations.  All these additional factors and technical challenges hindered the onset of the solar ALMA observing and needed to be addressed. To this end, two Solar Development studies were supported by the National Science Foundation (NSF), ESO and East Asia, and two corresponding teams, led by T.~Bastian (North America) and R.~Brajsa (ESO), were assembled in 2014. The aim of the international teams was to develop, implement and test solar observing modes, that could finally enable solar observations with ALMA. Building on the earlier work performed by the Joint ALMA Observatory in Chile, and colleagues in Europe, North America, and East Asia, the teams carried out several solar test campaigns as part of Commissioning and Science Verification/Extension of Capabilities (CSV/EOC) activities. The solution to the problem of receiver saturation,  found in ''de-tuning'' or ''de-biasing'' the ALMA superconductor-insulator-superconductor (SIS) mixers, as proposed by \citet{Yagoubov}, was thoroughly tested and introduced as a new operational mode referred to as the SIS mixer de-biased (MD) mode. The final (sixth) testing campaign, called ''a dress rehearsal'' for solar scientific observations, was conducted in December 2015. In connection with the ALMA development studies, the Solar Simulations for the Atacama Large Millimeter Observatory Network (SSALMON) was initiated, which supported the preparation for solar ALMA by the comprehensive science simulations effort with the contribution from many solar scientists \citep{Wedemeyer2015}. 

The outcome of these teams was in releasing of solar observing modes to the scientific community in the Cycle 4 proposal cycle \citep[October 2016 -- September 2017,][]{Andreani2016}. ALMA solar observations in Cycle 4 became possible owing to the dedicated work of the following colleagues \citep{Shimojo2016}:  A. Hales, A. Hirota, N. Philips, T. Sawada, I. de Gregorio-Montalvo, S. Corder (Joint ALMA Observatory); M. Shimojo, K. Iwai, S. Kim, S. Asayama, M. Sugimoto, K. Nakanishi, M. Saito, S. Iguchi (East Asia); T. Bastian, A. Remijan, S. White (North America); R. Brajsa, M. Barta, I. Skokic, P. Yagoubov, R. E. Hills (Europe).   

The solar observing modes were released as ''non-standard'' requiring  resorces additional to the ALMA data reduction pipeline, and were restricted to continuum observations in ALMA Bands 3 ($\lambda=3$~mm) and Band 6 ($\lambda=1.3$~mm) with time cadence of 2~sec. With the current design, water vapour radiometers (WVRs), which are used on ALMA antennas for correcting differential phase variations introduced by the overlying atmosphere at every antenna location, saturate when ALMA points to the Sun, and therefore WVR measurements are disabled for solar observations. Since these measurements are particularly important for long interferometric baselines, solar observations are constrained to the periods when the 12-meter array is in one of the three most compact configurations (with maximum baselines less than 500~m). The best spatial resolution offered in these compact configurations is 1.6\arcsec\ and 0.6\arcsec\ at $\lambda=3$~mm and $\lambda=1.3$~mm, respectively. For solar imaging the use of both the 7-meter array and 12-meter array simultaneous observations is essential to provide sufficent coverage of the spatial frequency domain (uv-plane).

The solar CSV/EOC activities are summarized in two solar ALMA reference papers by \citet{Shimojo2017a} and \citet{White2017a}, and are reviewed in \citet{White2017b} and \citet{Bastian2018}. The ALMA reference papers summarize the development and science-verification efforts leading up to the release of solar observing modes, provide the details of interferometric and single-dish mapping procedures, and give examples of observational sequences. The authors describe in detail the calibration of  the data, including gain, flux, and bandpass calibrations for interferometric data and absolute brightness calibration of single-dish observations. 

\citet{White2017a} report a new technique, Single Dish (SD) Fast Scan Mapping, developed and utilized for solar observing with ALMA during the commissioning phase (initially developed by Richard Hills and Neil Phillips for non-solar observing), which represents mapping of the full solar disk or of a smaller region with TP antennas using double-circle or Lissajous curve mapping pattern. Using this technique, maps of the full Sun (TP maps, 2400\arcsec x2400\arcsec) are generated within 4-20~min depending on wavelength. The excellent precision of the ALMA dishes leads to the high-quality fast-scan images (Figure~\ref{fig2}). In interferometric imaging such TP maps are used to fill in the short-spacing flux of the Sun. Though currently fast-scan mapping is available for the full Sun only, in the future this technique can be exploited for sub regions on the solar disk to study time-varying phenomena. The ALMA disk-center brightness temperatures, derived from the SD images by \citet{White2017a}, are 7300~K in Band~3 ($\lambda=3$~mm) and 5900~K in Band~6 ($\lambda=1.3$~mm), respectively, which is consistent with previous QS measurements at these wavelengths. The variations of brightness temperature on the disk were found to reach 2000~K at the 25\arcsec\ resolution of the data at 1.3~mm, with prominent contrast between quiet-Sun network and internetwok defining the variability seen on small spatial scales.

\begin{figure}
\label{fig3}
\begin{center}
\includegraphics*[width=14cm]{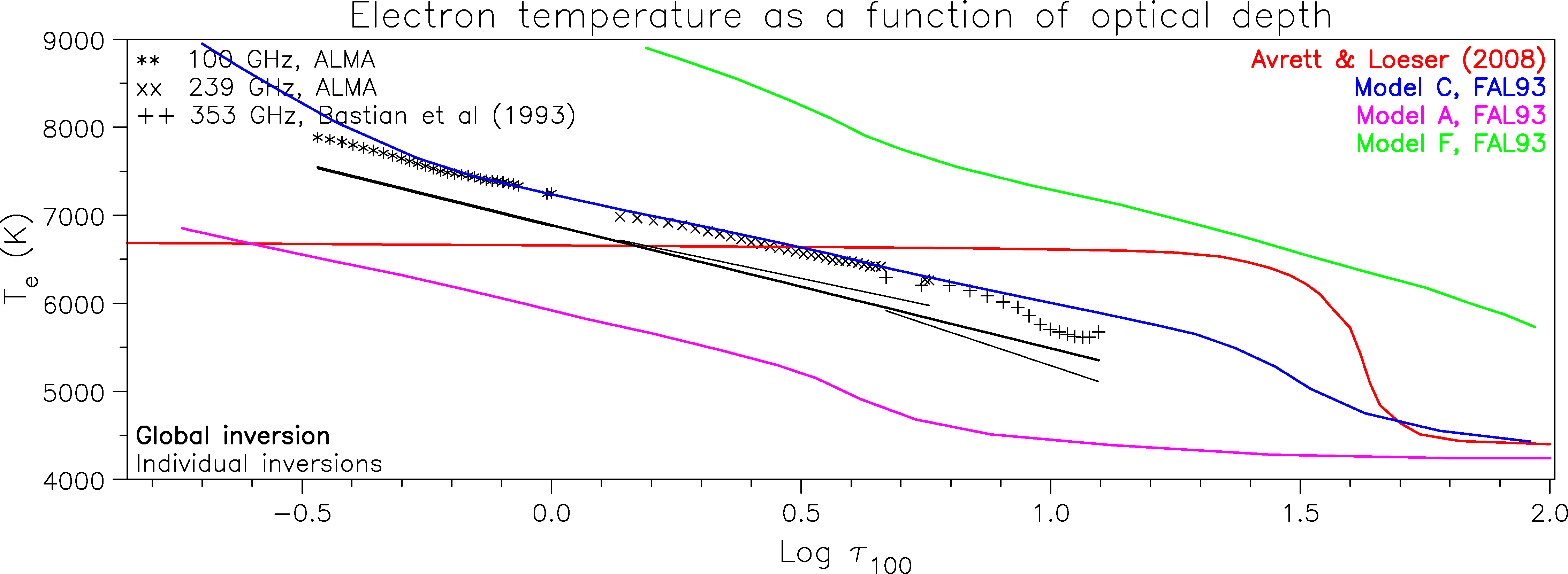}
\end{center}
\caption{The electron temperature as a function of the optical depth at $\lambda=3$~mm. The result of the inversion of the CLV curves is plotted with thick black line,  inversions of individual wavelengths are given with thin black lines. The symbols mark the Eddington-Barbier solution of the transfer equation. The temperatures computed from classical models are shown with colored lines.  Credit: \citet{aliss2017},  reproduced with permission \textcopyright ESO. }
\end{figure}

\section{First solar ALMA science}

The first solar ALMA observational data, including both the interfrometric regional images and superb fast-scan TP maps of the full disc, obtained during the solar science verification campaign, were thoroughly exploited by the solar community and a number of studies have been published. 

The first systematic comparison between various features of full-disk ALMA images from the SD observations at $\lambda=1.25$ and 3~mm and AIA/SDO images was reported in \citet{aliss2017}. To study center-to-limb (CLV) variations of the solar disk brightness from the ALMA images, the authors performed an inversion of the formal solution of the transfer equation for radio wavelengths, and obtained direct information on the variation of the electron temperature with the optical depth. The results of the inversion of  brightness CLV measurements (Figure~\ref{fig3}) were found to be consistent with a linear variation of electron temperature with the logarithm of the optical depth in the chromosphere and were only 5\% below the temperature from the classical atmospheric model by \citet{fontenla} representing the average quiet Sun. 

Fast-scan SD observations of the full Sun were also used in \citet{bra2018} to study various brightness structures seen in the ALMA image at $\lambda=1.2$~mm. By comparing ALMA brightness with full-disk solar images in H$\alpha$ line, in He I 1083 nm line core, with the AIA/SDO images at 170 nm, 30.4 nm, 21.1 nm, 19.3 nm, and 17.1 nm, and HMI/SDO magnetograms, the authors identified and described the mm-$\lambda$ brightness structures related to the quiet Sun, active regions, prominences on the disk, magnetic inversion lines, coronal holes and coronal bright points.   

\begin{figure}
\label{fig4}
\begin{center}
\includegraphics*[width=14cm]{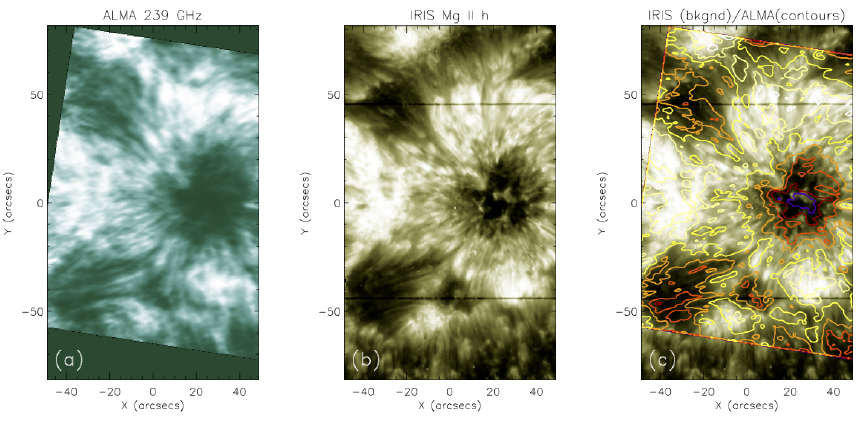}
\end{center}
\caption{Comparison between (a) the ALMA map at 1.25~mm and  (b) a map of IRIS Mg II h2v radiation temperature. (c) the contours of the 1.25~mm brightness overlaid on the IRIS radiation temperature map background.  From \citet{Bastian2017}.  \textcopyright AAS. Reproduced with permission.}
\end{figure}

\citet{Bastian2017} compared ALMA observations of the 1.25~mm-$\lambda$ continuum emission from the solar active region AR 12470 on 2015 December 18 with nearly simultaneous observations of the Mg II line intensities from the Interface Region Imaging Spectrograph (IRIS), which are believed to form at similar heights in the chromosphere (Figire~\ref{fig4}). Apart from the clear similarities between the emisions at the two wavelengths, distinct differences between the mm brightness temperature and the UV radiation temperature were found, including a compressed range of radiation temperatures of Mg II lines as compared to the brightness temperatures at 1.25~mm and an offset between the temperatures at two wavelengths.

\begin{figure}
\label{fig5}
\begin{center}
\includegraphics*[width=15cm]{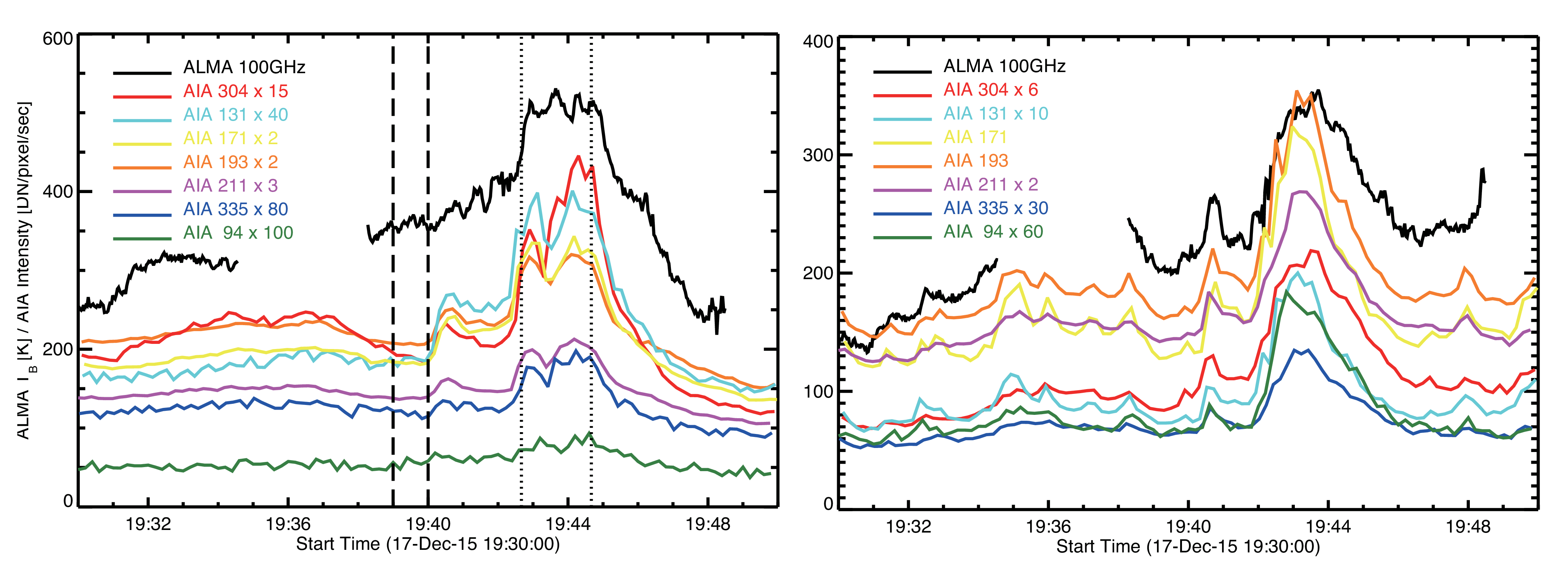}
\end{center}
\caption{Time profiles of the XBP and plasmoid intensity in the AIA/SDO EUV bands (colored lines) and from the ALMA data at 3~mm (black). Left: the total intensity from the area corresponding to the motion of the plasmoid. Right: the total intensity from the integrated area of the (stationary) XBP. Adapted from \citet{Shimojo2017b}.   \textcopyright AAS. Reproduced with permission.}
\end{figure}

Another study that made use of the science verification interferometric data was conducted by \citet{Shimojo2017b}. The authors reported the first ALMA observation of a solar plasmoid ejection from an X-ray bright point (XBP). By combining the radio (ALMA at $\lambda=3$~mm), EUV (AIA/SDO)  and X-ray (Hinode/XRT) data, the authors studied the thermal structure of the plasmoid. The intensity enhancements of the plasmoid at 3~mm  and in the EUV bands of AIA/SDO (Figure~\ref{fig5}) were attributed to either isothermal $10^5$ K plasma that is optically thin at 3~mm, or an optically thick $10^4$ K core with a hot envelope.  \citet{Shimojo2017b} demonstrated the value of the additional temperature and density constraints that ALMA provides for multi-wavelength studies of solar phenomena.

\begin{figure}
\label{fig6}
\begin{center}
\includegraphics*[width=14cm]{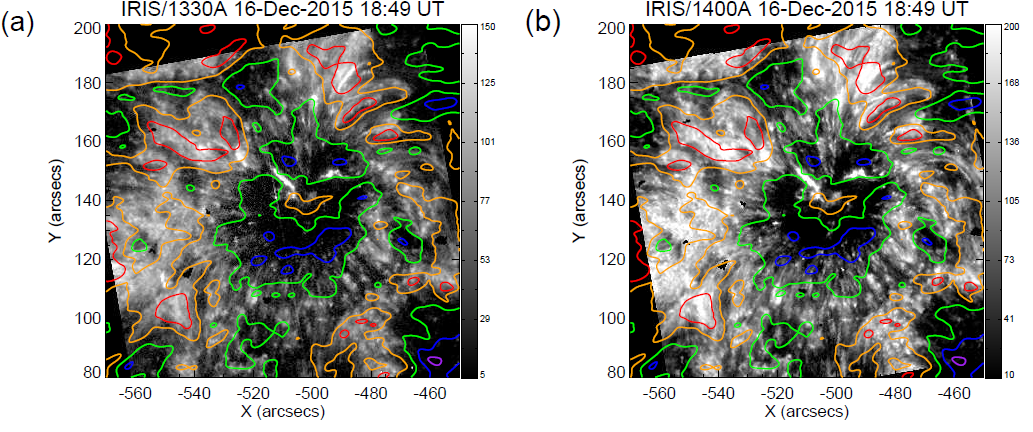}
\end{center}
\caption{IRIS slit-jaw images of AR12470 at (a) 1330\AA\ and (b) 1400\AA, with the overlaid color contours corresponding to the brightness temperature levels in the 3~mm-$\lambda$ map at 6300 K (purple), 6900 K (blue), 7500 K (green), 8100 K (orange), and 8700 K (red). From \citet{iwai2017}.  \textcopyright AAS. Reproduced with permission.}
\end{figure}

\citet{iwai2017} reported the discovery of a brightness enhancement in the center of a large sunspot umbra in the active region AR12470 at a wavelength of 3 mm, which was observed on 2015 December 16 in the mosaic mode with the spatial resolution of 4.9\arcsec x 2.2\arcsec. For the first time mm-$\lambda$ observations of sunspots have resolved umbral brightness structure at the chromospheric heights. Previous sunspot observations at mm wavelengths with the Berkeley-Illinois-Maryland Array (BIMA) and the Nobeyama Radio Heliograph (NoRH) were carried out with a resolution of around 10\arcsec, which was insufficient to clearly resolve the umbra within the sunspot.  Thus, in the BIMA maps at 3.5~mm \citep{Loukitcheva2014} the umbra was seen as the darkest feature in the image. Even at longer wavelengths, in the observations with the NoRH at $\lambda=8$~mm,  umbra was reported to be not brighter than the QS \citep{iwai2016}.  In the ALMA maps the enhanced emission at 3~mm was found to correspond to a temperature excess of $\sim$800~K relative to the surrounding penumbral region and to be nearly coincident with excess brightness in the IRIS 1330 and 1400\AA\ slit-jaw images, adjacent to a partial lightbridge (Figure~\ref{fig6}). The reported  brightness enhancement at 3~mm may be an intrinsic feature of the sunspot umbra at chromospheric heights, or it may be related to a coronal plume, or it also might be a manifestation of dynamic umbral flashes. Unfortunately, present observations cannot distinguish between different scenarios for the brightness enhancement, and additional ALMA observations are required to understand it.

\begin{figure}
\label{fig7}
\begin{center}
\includegraphics*[width=15cm]{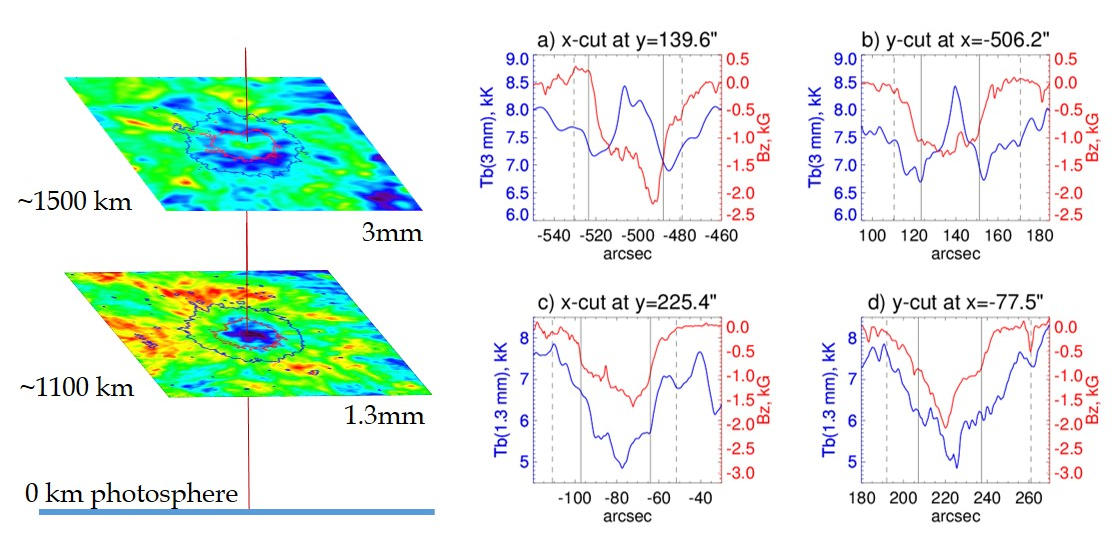}
\end{center}
\caption{ Left panel: ALMA images of AR12470 at 3 and 1.3mm.  Right panel: Profiles of the mm brightness (blue, left axis) and of the longitudinal component of the magnetic field from HMI/SDO (red, right axis) for (a) the x-cut at $y=139.6$\arcsec\, (b) the y-cut at $x=−506.2$\arcsec\ of the 3 mm-$\lambda$ image, (c) the x-cut at $y=225.4$\arcsec\, (d) the y-cut at $x=−77.5$\arcsec\ of the 1.3 mm-$\lambda$ image. The solid and dashed black lines indicate the positions of umbral and penumbral boundaries along the cuts, respectively.}
\end{figure}

In \citet{Loukitcheva2017} the study of \citet{iwai2017} was extended by analyzing the observations of the same sunspot at a wavelength of 1.3~mm, and by comparing ALMA observations at the two wavelengths with the predictions of sunspot umbral, and for the first time, penumbral models. The umbra was found to look very different at $\lambda=1.3$~mm from its image at 3~mm. In contrast to the brightness enhancement  observed  at 3~mm, at $\lambda=1.3$~mm the central part of umbra shows a temperature depression of $\sim$700~K relative to the QS level (Figure~\ref{fig7}). ALMA sunspot observations have also resolved the penumbral structure at mm wavelengths. In the ALMA images at 1.3~mm the penumbra is brighter than the QS, while at 3~mm its inner part is cooler than the QS. At both wavelengths the penumbral brightness ncreases towards the outer boundary (Figure~\ref{fig7}). The measured ALMA brightness temperatures for different parts of umbra and penumbra were compared with the expected mm-$\lambda$ brightness calculated from a number of standard and recently developed sunspot models, which differ from the reference QS model of \citet{fontenla} and from each other in the depth and extension of the temperature minimum region and in the location of the transition region. Among the tested umbral models that of \citet{Severino} provided the best fit to the observational data: the chromospheric temperature gradient in the model was found to be in reasonable agreement with the ALMA observations. According to this model, the bulk of the emission at $\lambda=1.3$ and 3~mm comes from the heights of $\sim$1100 km and 1500 km in the umbral chromosphere, respectively. No penumbral model gave a really satisfactory fit to the currently available measurements. It was concluded that multi-wavelength ALMA data can be used to validate or rule out models of the umbral atmosphere, and can also serve as an important constraint on any new empirical models. 

The results obtained from the first solar observations with ALMA persuasively demonstrated the diversity of potential solar research to expect from regular solar ALMA observing, started in Cycle 4. 

\section{ALMA Cycle 4 and beyond}

Over 50 solar proposals (3.6\% of the total number) were submitted for the first ALMA solar observing in Cycle 4. Fithteen proposals (3\% of the total number of the accepted proposals) were approved and allocated observing time. The solar observations were carried out in ''campaign'' mode, when the bulk of the ALMA solar observing was done within a fixed time window, in December 2016 and April 2017, in coordination with the space-borne instruments (IRIS and Hinode), as well as with a number of ground-based optical telescopes.  Due to novelty of the solar observing regime and very unstable weather conditions at the ALMA site during the solar observing campaign, Cycle 4 solar programs reached different stages of completion, from totally failed and partially completed programs to full completion and satisfactory data collection. After the data underwent calibration and quality assurance processes, the successfully carried out observations were delivered to their principal investigators (PIs). 
The standard ALMA data products include calibrated data in the form of the interferometric visibilities, scripts for image synthesis, and reference images (typically a time-averaged image).  Currently the work on imaging and analysis of the Cycle 4 data is being done. The data have a proprietary period of one year, after which they become available in the ALMA data archive\footnote{\url{https://almascience.nrao.edu/alma-data/archive}}. The overall outcome of the solar observations in Cycle 4 will be finalized after the studies based on the Cycle 4 data are published. But regardless of this, Cycle 4 is an important step towards establishing ALMA as a new, powerful and reliable solar observational instrument. 

The capabilities offered in ALMA Cycle 5 \citep[October 2017 -- September 2018,][]{Andreani2017}  were essentially the same as in Cycle 4, meaning interferometric and TP continuum observations in Bands 3 and 6, with time cadence reduced to 1~sec. Although there was a significant decrease in the number of solar proposals submitted  for Cycle 5, as compared to Cycle 4 (36 vs. 53), the amount of the proposals that received high-priority grades was similar to Cycle 4 (16, with a success rate of 44\%). In Cycle 5, solar observing is still considered as a ''non-standard' mode, but is no longer done in a ''campaign'' mode, meaning less flexibilty in scheduling than in Cycle 4 due to  priority scheduling of high-ranked non-solar day time programs. Therefore, the throughput in the number of solar programs executed so far in Cycle 5 is lower than it was in Cycle 4. Thirty two solar proposals submitted for Cycle 6 observations are currently under review and the results of the proposal call will be announced in July 2018.

Currently, new capabilities and observing modes for solar ALMA are undergoing testing and commissioning \citep{Bastian2018}. The Solar Development team intends to provide support for observations in submillimeter frequency bands, Band 7 (0.8--1.0~mm) and possibly Band 9 (0.4--0.5~mm), starting from Cycle 7. This will allow to study the layers of the lower chromosphere down to the temperature minimum region, and thus to expand the potential of ALMA based chromospheric diagnostics. Among the other objectives for testing and comissioning for Cycle 7 are support for spectral-line-mode observing and subregional mapping on a cadence of tens of seconds. Testing and comissioning of ALMA capability of solar circular polarization measurements, which are of fundamental importance for diagnostic of  chromospheric magnetic field, is also a high-priority task in the Cycle 7 testing plans of the Solar Development team. Among the capabilities awaiting future ALMA cycles is support for simultaneous multiband (multi-wavelength) observing, which implies dividing the array into several subarrays that are each capable of observing the Sun independently. In contrast to the SIS mixer de-biased mode employed for observing the non-flaring Sun, observations of solar flares require the use of solar filters, which has been previously demonstrated but needs further development and testing. And finally, an updated design of the water vapour radiometers is required to remove the restriction of using only compact array configurations for solar observing.

\section{Conсlusions}

ALMA is a unique instrument that is capable of providing the necessary spatial, temporal and spectral resolution to explore the central questions in modern solar physics. The first solar observations obtained during the comissioning and science verification phase have verified ALMA's potential in this respect. Owing to the unique experience and extensive effort of solar radio astronomy enthusiasts, regular solar ALMA observations have finally became possible. Cycle 4 was the first attempt to observe the Sun with ALMA under regular conditions and it provided the community with a wealth of chromospheric data to be analyzed. While current capabilities of solar ALMA observations remain limited, they represent a major advance over observational capabilities previously available at millimeter wavelengths and will be further improved in the nearest future.  In summary, the era of ALMA in solar physics has finally started.

\section{Acknowlegment}

ALMA is a partnership of ESO (representing its member states), NSF (USA) and NINS (Japan), together with NRC (Canada), NSC and ASIAA (Taiwan), and KASI (Republic of South Korea), in cooperation with the Republic of Chile. The Joint ALMA Observatory is operated by ESO, AUI/NRAO and NAOJ. This work has been supported by Russian RFBR grant 16-02-00749.

\end{document}